\documentclass[doublecol]{epl2} 
\usepackage{graphicx}
\usepackage{amssymb}
\usepackage{amsmath}
\usepackage{amsfonts}
\renewcommand{\phi}{\varphi}

\title{From single-particle to collective effective 
temperatures in an active fluid of self-propelled particles}

\shorttitle{Effective temperatures in active fluids}

\author{Demian Levis \inst{1} \and Ludovic Berthier \inst{2}}

\institute{
\inst{1} Departament de F\'isica Fonamental, Universitat de Barcelona, 
Av. Diagonal 645, E08028 Barcelona, Spain \\
\inst{2} Laboratoire Charles Coulomb, UMR 5221, CNRS and Universit\'e
Montpellier, Montpellier, France}

\pacs{05.10.Ln}{Monte-Carlo methods}
\pacs{47.57.-s}{Complex fluids and colloidal systems}

\abstract{We present a comprehensive analysis of
effective temperatures based on fluctuation-dissipation relations
in a model of an active fluid composed of 
self-propelled hard disks. We first investigate the relevance 
of effective temperatures in the dilute and moderately dense 
fluids. We find that a unique effective temperature does not
in general characterize the non-equilibrium dynamics of
the active fluid over this broad range of densities, because
fluctuation-dissipation relations yield a lengthscale-dependent  
effective temperature. By contrast, we find that 
the approach to a non-equilibrium glass transition 
at very large densities is accompanied by the emergence of a unique 
effective temperature shared by fluctuations at all lengthscales. This suggests 
that an effective thermal dynamics generically emerges at long times 
in very dense suspensions of active particles due to 
the collective freezing occurring at non-equilibrium glass transitions.}
 
\begin{document}

\maketitle

\section{Introduction}
\label{sec:intro}

Statistical mechanics provides a unified theoretical description of systems 
at thermal equilibrium in terms of the probability distribution over 
phase space, from which thermodynamic quantities such 
as temperature can be defined~\cite{ChandlerBook}. 
A similar framework is lacking for out-of-equilibrium systems
for which the definition of a temperature remains an open 
issue~\cite{Kurchan2005}. 
Active matter formed by assemblies of living cells~\cite{Angelini2011}, 
bacteria~\cite{WuLibchaber2000}, self-propelled 
colloids~\cite{Palacci2010,Theurkauff2012,Buttinoni2013,Ginot2014} 
or grains~\cite{Deseigne2010,ramaswamy}, 
is a coherent class of non-equilibrium 
systems receiving increasing attention, 
because they raise fundamental issues and for
potential applications in soft matter and 
biophysics~\cite{Marchetti2012a, VicsekReview}. 
A number of recent studies have addressed the question of whether 
``effective'' thermodynamic concepts can be fruitfully applied to 
describe the phase behaviour and microscopic dynamics 
of active matter. This question is natural because if some mapping
to an equilibrium situation exists, then the whole arsenal 
of equilibrium statistical mechanics becomes available for further
theoretical treatment. In particular the definition of a non-equilibrium 
temperature~\cite{Loi2008,Wang2011a,Loi2011,Ben-Isaac2011,Szamel2014a,Suma2014}, 
of an active pressure~\cite{Ginot2014,Takatori2014,Solon2014}, 
of activity-induced interactions~\cite{Ginot2014,Farage2015} 
and non-equilibrium free energies~\cite{Tailleur2008,Speck2014} 
have been investigated 
for self-propelled particles. 

An effective temperature $T_{\rm eff}$ defined as the parameter 
replacing the thermal bath temperature in
fluctuation-dissipation relations was extensively 
studied in slowly relaxing materials, such as
spin and structural 
glasses~\cite{Cugliandolo1993,Cugliandolo1994,Crisanti2003,Cugliandolo2011}. 
In these systems, 
the effective temperature becomes a meaningful thermodynamic 
concept~\cite{Cugliandolo1997}. 
It can be used to quantify heat flows and can be defined through 
a microcanonical construction over a restricted region of the phase space.
Physically, this approach is possible because there is a
strong decoupling of timescales~\cite{Kurchan2005} 
between thermal motion at short times,
which essentially follows an equilibrium statistics, and an
``effective'' thermal dynamics at long times, which is an emerging 
collective property characterizing slow dynamical events 
leading to structural relaxation in driven and aging glasses.

In active matter, the possibility to define a {\it single-particle}  
effective temperature in the dilute regime where particles do not
interact has been 
discussed~\cite{Tailleur2009,Ben-Isaac2011,Wang2011a,Szamel2014a,Maggi2014}.
The simple fluid state has been analysed mostly by numerical 
simulations determining fluctuation-dissipation relations 
for different observables in various models of self-propelled particles 
\cite{Loi2008,Loi2011,Suma2014}. Finally, in the limit of 
very large densities where 
self-propelled particles may undergo a non-equilibrium 
glass transition, the driven glassy dynamics should 
resemble the one of slowly driven glasses and 
a {\it collective} effective temperature
was predicted to emerge from the analysis of a mean-field active 
glass model~\cite{Berthier2013}, but this prediction
has not been tested in a realistic model in finite dimension. 
In this work, we focus on a model of an active fluid composed of 
self-propelled hard disks and analyse fluctuation-dissipation 
relations over a very broad range of densities encompassing the dilute,
fluid and glassy regimes, in order to provide a comprehensive 
picture of the relevance of effective temperatures in active matter. 
In particular, we find 
strong violations to the fluctuation-dissipation relations 
in all regimes, but conclude that effective temperatures 
appear most relevant in the slowly relaxing glassy regime. 

\section{Numerical model}

We consider a two-dimensional system of $N$ 
interacting self-propelled hard disks of diameter $\sigma$, enclosed in 
a $L\times L$ square box with periodic boundary conditions.  Self-propulsion 
is modelled by a non-Markovian stochastic drive, implemented by a kinetic 
Monte-Carlo (MC) rule that we describe below. The phase 
behaviour of the model and its microscopic dynamics were investigated 
before~\cite{Levis2014,Berthier2014,Ginot2014}. Here we extend these
studies to analyse fluctuation-dissipation relations and effective 
temperatures over a large range of densities.

A particle configuration is described by $\{ \boldsymbol{r}_i(t), i 
=1 \cdots N \}$, the set of  
particle positions at time $t$. The dynamics proceeds as follows. 
At time $t$, a particle $i$ is chosen at random and a small displacement 
$\boldsymbol{\delta}_i(t)$ of 
amplitude  $|\boldsymbol{\delta}_i(t)|<\delta_0$ 
is proposed. Just as for equilibrium hard disks~\cite{KrauthBook}, the
particle position is updated according to
$\boldsymbol{r}_i(t+\Delta t)=\boldsymbol{r}_i(t)+\boldsymbol{\delta}_i(t)
P_{\text{hard}}$, where $P_{\text{hard}}=1$ if the move does not 
generate any overlap with a neighbouring disk, 
and $P_{\text{hard}}=0$ otherwise. 

Self-propulsion is introduced via a finite persistence time 
of the successive moves $\boldsymbol{\delta}_i(t)$. This non-Markovian
dynamics breaks 
detailed balance. In practice, we choose
$\boldsymbol{\delta}_i(t+\Delta t)=\boldsymbol{\delta}_i(t)+
\boldsymbol{\eta}_i(t)$,
where $\boldsymbol{\eta}_i(t)$ is a uniformly distributed random shift of 
amplitude 
$|\boldsymbol{\eta}_i(t)|<\delta_1$,  chosen independently at each step. 
As a consequence, successive displacements $\boldsymbol{\delta}_i$ 
decorrelate after a persistence time $\tau=(\delta_0/\delta_1)^2$, 
and this dynamics generates persistent random walks
for isolated particles. (See Ref.~\cite{Levis2014} for more details about
the model.)  
This model is appealing because it has only two control parameters.
The persistence time $\tau$ controls the self-propulsion of the particles, 
whereas the packing fraction $\phi=\frac{\pi N \sigma^2}{4L^2}$
quantifies excluded volume effects. In addition, the equilibrium 
hard disk model, where
the physics is uniquely controlled by $\phi$, is restored in 
the limit $\tau \to 0$. The model does not include more 
complex features such as alignment rules or hydrodynamic interactions
and serves as a minimal model to study the direct competition
between glassiness and self-propulsion. 
A continuous-time version of the model for arbitrary particle interactions 
has recently appeared~\cite{Szamel2015}.

We perform simulations using $N=10^3$ particles, varying $\phi$ and 
$\tau$. We vary $\phi$ from the dilute limit $\phi \to 0$ 
up to $\phi=0.825$, well 
beyond the equilibrium glass transition of hard disks $\phi^{\rm eq}_c
\approx 0.80$~\cite{Donev2006}, 
and $\tau$ in the range $\tau \in [0, 10^4 ]$. 
We fix for convenience $\delta_0=\sigma/10$. 
From now on, we set the Boltzmann constant $k_B=1$, we use $\sigma$ as 
the unit of length and one MC step represents $N$ attempted moves. 
To suppress crystallisation in the glassy regime, we 
use a 35:65 binary mixture of hard disks with diameter ratio 
$\sigma_1/\sigma_2=1.4$ ($\sigma_1$ then being the unit length) for systems 
above $\phi=0.69 $. For lower $\phi$, all systems 
are monodisperse. Size polydispersity is introduced for convenience, 
but it does not influence the results presented below. 

\section{Fluctuation-dissipation relations}

Using linear response theory at equilibrium, one can prove that
the response of a system to an infinitesimal external perturbation 
is related to the spontaneous fluctuations of its conjugate observable
via the fluctuation-dissipation theorem (FDT)~\cite{ChandlerBook}. 
Consider a system perturbed at 
time $t=0$ by an external field of constant amplitude $f_0$
coupled to an observable $B$. We define the cross-correlation function
between observables $A(t)$ and $B(t)$ as 
$C_{AB}(t,t^{\prime})=\langle A(t)B(t^{\prime})\rangle_0-\langle 
A(t)\rangle_0 \langle B(t^{\prime})\rangle_0$, where 
$\langle ... \rangle_0$ denotes an equilibrium ensemble average.
The conjugate linear response function $R_{AB}(t)$ is defined as 
\begin{equation}
\langle A(t) \rangle - \langle A(t) \rangle_0= f_0 \int_{0}^t\, 
R_{AB}(t,t^{\prime})\text{d}t^{\prime}+ {\cal O}(f_0^2),
\end{equation}
where $\langle ... \rangle$ represents an average in the presence of the 
perturbation.
The time integral, $\chi_{AB}(t,t')=\int_{t'}^t R_{AB}(t,u) 
\text{d}u$, is the linear susceptibility, which is a more 
easily accessible quantity than $R_{AB}$ both in experiments and simulations.
In terms of $\chi_{AB}$, the FDT becomes a simple linear relation
\begin{equation}\label{eq:FDT}
\chi_{AB}(t,t')=\frac{1}{T} \left[C_{AB}(t,t)-C_{AB}(t,t')\right]\ .
\end{equation}

Far from equilibrium, the FDT has no reason to be obeyed.
However, it has been shown that in systems with slow dynamics,
like spin glasses or supercooled liquids, the linear relation 
eq.~(\ref{eq:FDT}) is verified 
by replacing $T$ by an \emph{effective temperature} 
$T_{\rm eff}$~\cite{Crisanti2003,Cugliandolo2011}. 
Technically, one introduces the fluctuation-dissipation ratio $X_{AB}(t,t')$ 
through~\cite{Cugliandolo1993}
\begin{equation}
R_{AB}(t,t')=\frac{X_{AB}(t,t')}{T}\frac{\partial}{\partial t^{\prime}}
C_{AB}(t,t^{\prime})  .
\end{equation}
If $X_{AB}(t,t')$ is slaved to the relaxation and becomes 
a function of the variable $C_{AB}(t,t')$, 
an effective temperature $T^{AB}_{\rm eff}(C)=T/X_{AB}(C)$ can then be defined 
from the slope of a ``fluctuation-dissipation (FD) plot'' 
of $\chi_{AB}$ vs. $C_{AB}$ parametrized by the time
difference $t-t'$. Such plots have repeatedly 
appeared in the literature of glassy materials. 
The equilibrium FDT is recovered when $X_{AB}(C)=1$ 
and the slope of the parametric FD plot is then simply $-1/T$. 
The effective temperature defined above may in general depend on the 
chosen dynamic observables $A$ and $B$ but 
in order to allow for a thermodynamic interpretation, $T_{\rm eff}$ should 
be independent of this choice. This is of course the case in 
equilibrium and this hypothesis has been tested extensively in the glass 
literature~\cite{Crisanti2003,Berthier2002,Sollich2002}. 

We will investigate whether this important 
property also holds in our model. To this end, we first 
analyse the effect of an external 
constant force $\boldsymbol{f}_i=\epsilon_i f_0\boldsymbol{e_x}$ applied 
from time $t=0$ in the $x$-direction, 
where $\epsilon_i = \pm 1$ with equal probability. 
In our simulations, the perturbation is introduced via the acceptance 
probability of a MC update,
\begin{equation} 
\label{eq:Pacc0}
P_{\text{acc}}(\boldsymbol{\delta}_i , f_0)=
\min\left[1,\, e^{\epsilon_if_0\delta x_i}\right] P_{\text{hard}} ,
\end{equation} 
where $\delta x_ i = \boldsymbol{\delta}_i \cdot \boldsymbol{e_x}$
and the equilibrium temperature is $T=1$.
For $f_0\neq 0$, this introduces a bias in the $x$-direction, 
whereas the motion in the transverse $y$-direction is unaffected. We 
carefully checked in all our simulations that the value of $f_0$ we use 
is small enough to probe the linear regime so that deviations 
from the equilibrium FDT relations are uniquely due to the 
non-equilibrium nature of the self-propelled dynamics.
The associated susceptibility is the displacement of the 
particles induced by $f_0$:
\begin{equation}
\chi_0(t)= \lim_{f_0\to 0} \frac{1}{N} 
\sum_{i=1}^N \frac{\epsilon_i}{f_0} {\langle x_i(t) -x_i(0)\rangle} .
\end{equation}
For these observables, the FDT in eq.~(\ref{eq:FDT}) reads:
\begin{equation}
\chi_0(t) = \frac{1}{2T} \Delta x^2(t),
\end{equation}
where $\Delta x^2(t)= N^{-1} \sum_{i=1}^N \langle\left( 
x_i(t)-x_i(0)\right)^2\rangle_0$ is the mean-squared displacement.  
In the long-time regime $\Delta x^2(t)\sim 2D t$ and $\chi_0(t)\sim \mu t$, 
where $D$ is the diffusion coefficient and $\mu$ the mobility. 
At long times, the FDT yields the Stokes-Einstein relation, 
which is generalized out-of-equilibrium to 
\begin{equation}\label{eq:StokesEinstein}
T_{\rm eff} = D/\mu,
\end{equation} 
with an effective temperature $T_{\rm eff}$ that in principle depends 
on both $\tau$ and $\phi$. 

In order to analyse carefully to what extend $T_{\rm eff}$ depends on 
the choice of observables, we consider also 
$A_q$ and $B_q$ defined as~\cite{Berthier2002}
\begin{equation}\label{eq:AB}
A_q(t)=\frac{1}{N} \sum_{j=1}^N \epsilon_j e^{i q x_j(t)},\
B_q(t)=2\sum_{j=1}^N \epsilon_j \cos[q x_j(t)], 
\end{equation}
where $x_j(t) = \boldsymbol{r}_j(t) \cdot \boldsymbol{e_x}$. 
With this choice, the associated correlation function entering in 
eq.~(\ref{eq:FDT}) is the self-intermediate scattering function
\begin{equation}
F_s(q,t) = \frac{1}{N}
\langle \sum_{j=1}^N e^{i q [x_j(t)-x_j(0)]}\rangle_0, 
\end{equation}
and the corresponding linear susceptibility is 
\begin{equation}
\chi_q(t)= \lim_{f_0 \to 0} {\langle A_q(t) - A_q(0) \rangle} /{f_0}\, .
\end{equation} 
This perturbation is physically meaningful, as its wavevector dependence
allows to probe the FDT at various lengthscales. 
Numerically, we use 
\begin{equation}
P_{\text{acc}}(\delta x_i,f_0)=\min\left[1, e^{2\epsilon_i f_0 
[\cos(q x_i^{\prime})-
\cos(q x_i)]}  \right]  P_{\text{hard}} ,
\end{equation}
where  $x_i^{\prime}=x_i+\delta x_i$.
A non-equilibrium effective temperature $T_{\rm eff} \neq 1$ 
is then revealed if a  
straight line appears in the FD plot $\chi_q$ vs. $[1-F_s(q,t)]$ for 
a given value of $q$. In the limit of small $q$, the long-time
diffusion is probed and these observables reduce to the 
Stokes-Einstein relation.

\section{Effective temperatures in the active fluid}

In the dilute limit $\phi\to 0$, the only control parameter 
is the persistence time $\tau$. In this regime, particles have ballistic 
motion at short times, $t < \tau$, which crosses
over to diffusive behaviour at long times, $t > \tau$, 
with $D \propto \tau$~\cite{Levis2014}.
The displacement $\chi_0$  induced by a constant force 
does not depend on the noise correlations, and hence is also independent 
of the persistence time. The mobility is then equal to the one 
of non-persistent Brownian particles. Thus, from the Stokes-Einstein
relation we find 
\begin{equation}\label{eq:StokesEinstein2}
T_{\rm eff} ( \tau, \phi\to 0) = D / \mu \propto \tau.
\end{equation} 
This linear dependence on the persistence time obviously crosses over to 
a constant $T_{\rm eff} \to 1$ as $\tau \to 0$ where equilibrium is recovered.
Similar results were obtained before in a variety 
of active models~\cite{Tailleur2009,Ben-Isaac2011,Szamel2014a}.

Using the sedimentation of self-propelled particles, a meaningful 
$T_{\rm eff}$ should also be accessible from the density 
profiles $\phi(z)$ along the direction of the gravity field. 
This has recently been 
investigated~\cite{Ginot2014} experimentally in suspensions of Janus 
colloids and numerically using the present model.
In the dilute limit, at the top of the sediment, both the 
the Janus suspension and the self-propelled hard disk model were found to  
display an exponential density profile,
just as an equilibrium ideal gas at temperature $T_{\rm eff}$. 
Up to numerical accuracy, the effective temperature extracted from this 
static measurement is equal to the one obtained from the 
Stokes-Einstein relation. Thus, dilute suspensions 
of self-propelled particles can be seen as 
a ``hot'' ideal gas with effective temperature $T_{\rm eff} \propto \tau$.
The agreement between sedimentation and Stokes-Einstein relation
is reasonable, as the gravity represents a biasing field acting
at $q=0$.
Overall, this suggests that a {\it single-particle} effective temperature 
is useful to characterize dilute active particles at large enough 
lengthscales, where details of the self-propulsion mechanism 
do not matter.
 
\begin{figure}
\centering
\includegraphics[width=8.5cm]{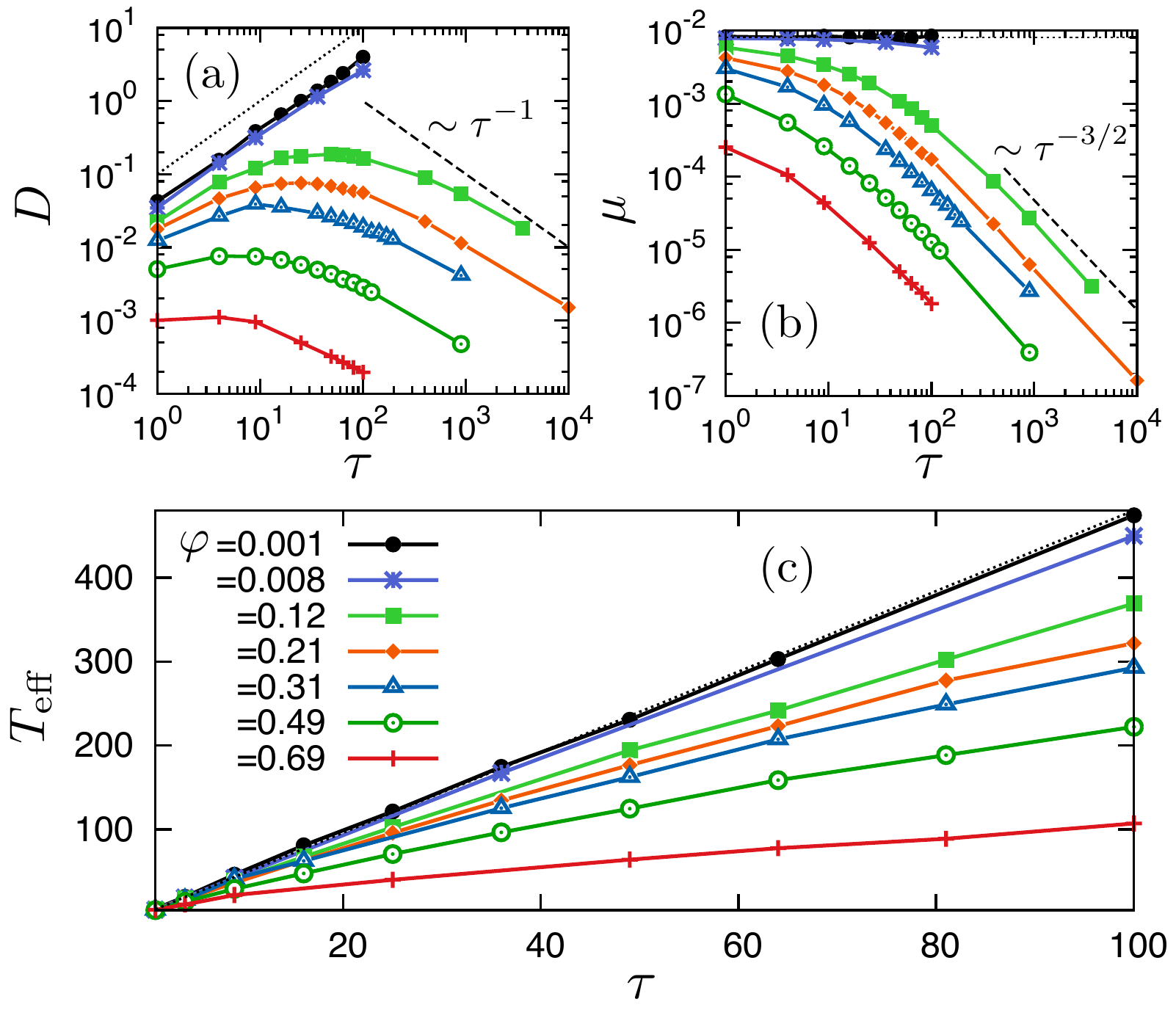}

\caption{Diffusivity (a), mobility (b) and 
effective temperature (c) as a function of  $\tau$ for 
several packing fractions in the active fluid. 
In the dilute regime, $D \sim \tau$,
$\mu \sim const$, $T_{\rm eff} \sim \sqrt{\tau}$ (dotted lines). 
In the clustering regime, $D \sim \tau^{-1}$,
$\mu \sim \tau^{-3/2}$, $T_{\rm eff} \sim \sqrt{\tau}$
(dashed lines in (a) and (b)). 
The effective temperature increases with $\tau$, but it 
decreases with $\phi$.} 
\label{fig:mu}
\end{figure}

We now turn to finite densities, where 
many-body effects may modify the simple behaviour found in  
eq.~(\ref{eq:StokesEinstein2}). 
In Fig.~\ref{fig:mu}(a-c), we show the results of numerical simulations
for $D$, $\mu$ and $T_{\rm eff}$ for a broad range of $\tau$ 
and $\phi$ values. When $\phi \to 0$, the leading 
behaviour exposed above is confirmed by the numerics. 
As discussed in detail in~\cite{Levis2014}, the diffusion 
coefficient $D$ at finite density is a non-monotonic function of 
the persistence time, and a density-dependent optimum value for a persistence 
time $\tau^*(\phi)$ appears where the diffusivity is maximal,
see Fig.~\ref{fig:mu}(a).
This maximum arises because self-propulsion 
simultaneously accelerates the dynamics of single particles, 
but it also leads to the formation of dynamic clusters 
where particles can be trapped. Physically, 
$\tau^*$ delimits the boundary between a homogenous fluid 
phase and a cluster phase characterised by the presence 
of fractal aggregates. Deep in the cluster phase, 
the diffusivity decreases as $D \sim \tau^{-1}$, because it is controlled
by the residence time of particles at the {\it surface} of the 
clusters~\cite{Levis2014}.

By contrast, our simulations indicate that 
the mobility decreases monotonically with increasing $\tau$
at constant density, see Fig.~\ref{fig:mu}(b). This 
simple behaviour arises because self-propulsion does not 
enhance the mobility of isolated particles but it 
strongly affects its value in the activity-induced cluster phase. 
Asymptotically, we find that $\mu \sim \tau^{-3/2}$,
suggesting that $\mu$ is dominated by 
the average time spent by the particles {\it inside} the clusters.
This behaviour can be understood 
by considering that particles either diffuse freely between 
the clusters with a diffusion constant given by 
$D(\phi\to 0) \propto \tau$, or are kinetically trapped 
within clusters of average size $n \propto \sqrt{\tau}$~\cite{Levis2014}.
The time spent inside the clusters thus scales as 
$\tau \times n$, so that the mobility scales as
$\tau^{-3/2}$, as observed numerically
in the large-$\tau$ limit.

By taking the ratio of $D$ and $\mu$, we obtain 
the $q \to 0$ effective temperature shown in 
Fig.~\ref{fig:mu}(c). The dilute limit behaviour 
$T_{\rm eff} \sim \tau$ is numerically recovered. At finite density, 
we observe that $T_{\rm eff}$ increases monotonically with 
$\tau$, and is thus not affected by the non-monotonic 
variation of the diffusion constant. However, 
the effect of the interactions at finite density
is obvious as the effective temperature grows
much more slowly with $\tau$ when $\phi >0$. In fact, the data
shows that, for a fixed value of the persistence time, 
$T_{\rm eff}$ decreases strongly with $\phi$, showing that 
interactions actually decrease the ``agitation'' 
of self-propelled particles. Indeed, in the strongly interacting 
regime, the combination of $D \sim 
\tau^{-1}$ and $\mu\sim\tau^{-3/2}$ suggests that 
$T_{\rm eff} (\tau,\phi) \sim \sqrt{\tau}$. Therefore, 
the scaling with persistence time {\it at finite density} is 
qualitatively different from the linear behaviour found in the 
dilute regime. 

In a model of repulsive active dumbbells studied 
recently in the fluid regime, $T_{\rm eff}$ has a 
non-monotonic dependence on the activity that becomes more pronounced 
by increasing the density~\cite{Suma2014}. 
In our model, many-body effects alter the scaling of $T_{\rm eff}$ with 
$\tau$ in a simpler manner, 
monotonically reducing the effective ``heating'' of the 
system.

\begin{figure}
\centering
\includegraphics[width=8.5cm]{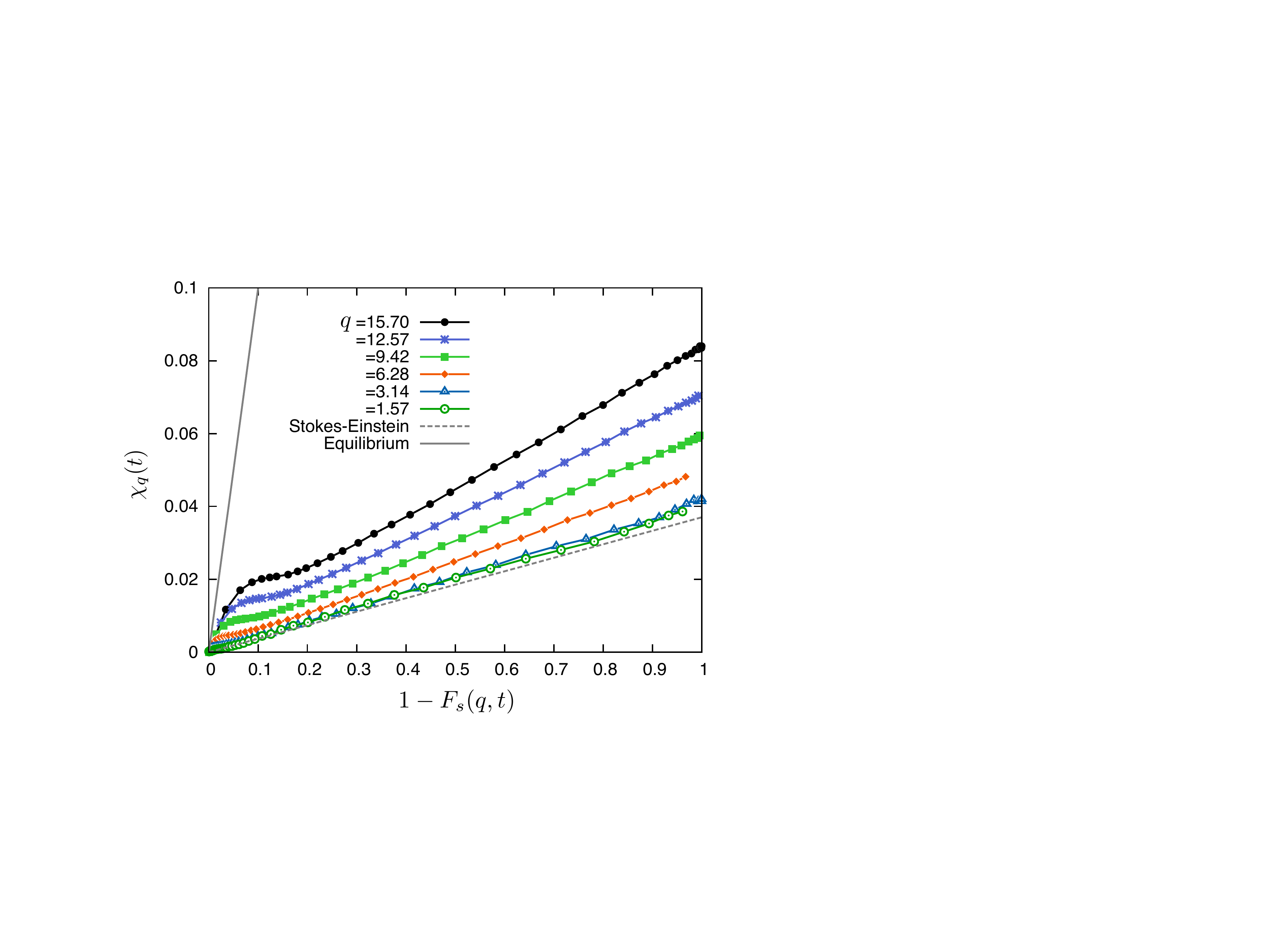}
\caption{Parametric FD plots obtained in the active fluid 
for $\tau= 9$ and $\phi=0.747$ for different values of the 
wavevector $q$ characterizing the perturbation. 
Whereas the Stokes-Einstein value for $T_{\rm eff}$ is recovered 
in the $q \to 0$ limit, a different value of  $T_{\rm eff}$
characterizes each $q$, suggesting that the active fluid regime 
is not characterized by a single effective tempetature. 
Equilibrium FDT is shown as a full line.}
\label{fig:PF747_vark}
\end{figure}

We now analyse perturbations at finite 
lengthscales, to probe more deeply the  
thermodynamic meaning of the effective 
temperature determined at large lengthscales
from the Stokes-Einstein relation.
In Fig.~\ref{fig:PF747_vark} we show 
representative parametric FD plots $\chi_q(t)$ 
versus $F_s(q,t)$ for our model at $\phi=0.747$ with a fixed $\tau = 9$ and 
several values of the wavevector $q$. 
As expected, the correlation and response 
functions of the system deviate strongly from the equilibrium FDT. 
However, the FD plots all follow a convincing straight line. 
This seems to imply that for each wavevector, 
a well-defined value of the effective temperature 
characterizes the long-time dynamics of the system.
However, we also observe that the value of this
effective temperature continuously depends on 
the chosen wavevector, and $T_{\rm eff}$ decreases when $q$ is increased. 
This behaviour suggests that a unique effective temperature does not 
describe the dynamics of our model in the active fluid regime. 

These observations extend to finite densities the results 
obtained analytically in the dilute regime of active Brownian particles, 
where it is similarly found that perturbations at finite 
lengthscales cannot be described in an effective 
thermodynamical framework \cite{Szamel2014a}. 
This conclusion therefore applies to the entire fluid regime, 
where $T_{\rm eff}$ sensitively depends on the choice of conjugate
observables. More broadly, these results echo recent 
studies showing that extensions of equilibrium concepts 
to active fluids is in many instances not straightforward
\cite{Tailleur2008,Tailleur2009,Redner2013,Fily2012,Solon2014,Speck2014}. 

\section{Collective effective temperature
in the glassy regime}

We finally investigate the model at very high densities, 
approaching dynamical arrest from the fluid. 
Recent work supports the idea that dense 
assemblies of self-propelled particles can display a dramatic slow down 
of the dynamics sharing strong analogies with the glassy dynamics of 
particle systems in contact with a thermal 
bath~\cite{BerthierKurchan2013,Ni2013,Berthier2014,Wysocki2014a,Brader2014,Szamel2015,dasgupta}. 
It was found in particular that activity shifts the glass 
transition towards higher densities (or lower temperatures), 
but leaves the main collective 
properties of the slow relaxation near the transition qualitatively 
unchanged~\cite{BerthierKurchan2013, Ni2013, Berthier2014}.  

\begin{figure}
\centering
\vspace{0.2cm}
\includegraphics[width=8.5cm]{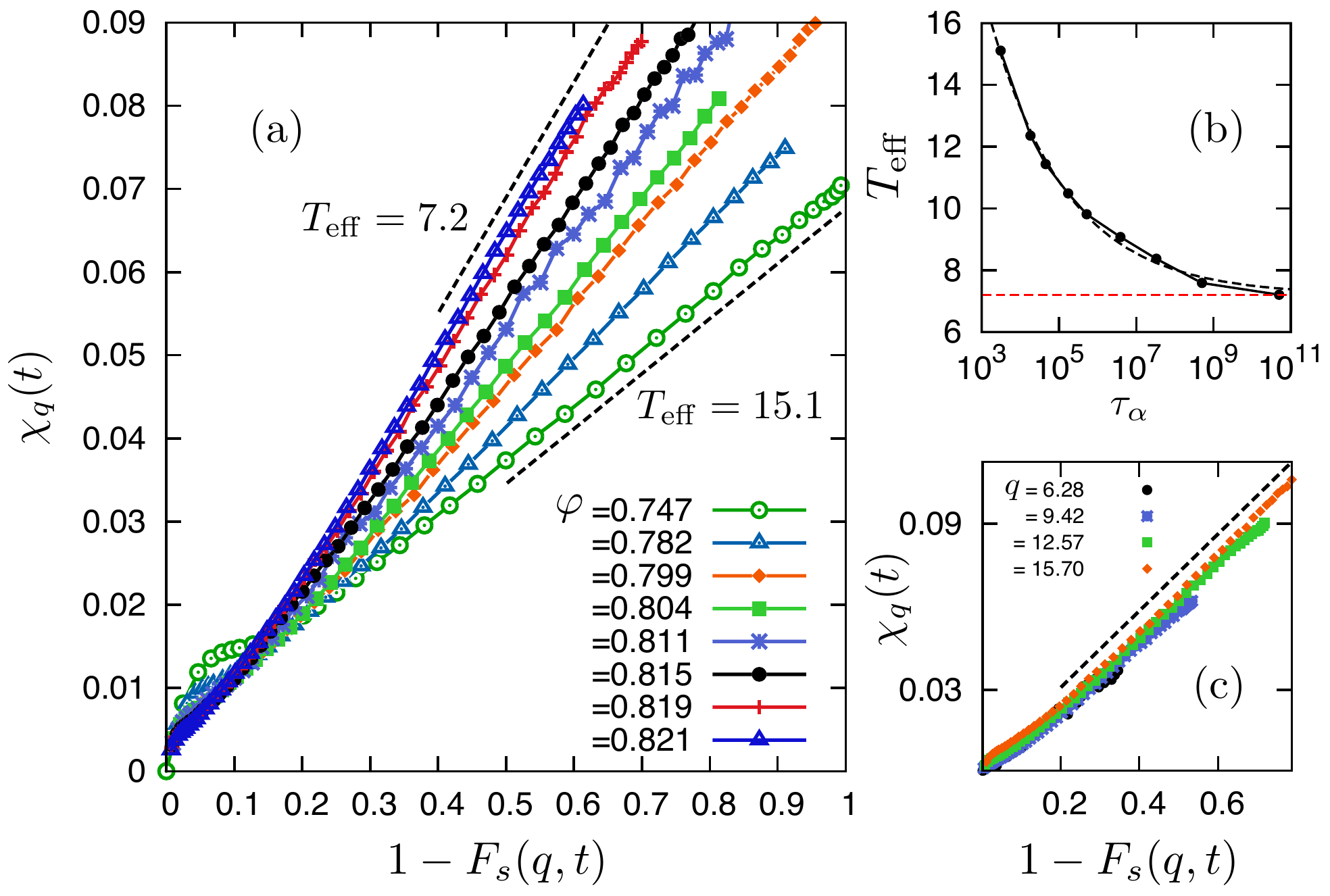}
\caption{(a): Parametric FD plots for $q= 4\pi$ and 
$\tau=9$ and different volume fractions $\phi$ 
approaching the non-equilibrium glass transition. 
(b) Effective temperatures extracted from 
(a) plotted as a function of the relaxation time $\tau_{\alpha}$. The 
horizontal dotted line corresponds to the estimated asymptotic value 
of $T_{\rm eff}=7.2$ as $\tau_\alpha \to \infty$.
The dashed line corresponds to 
an empirical power law fit, 
$T_{\rm eff} = 7.2 + 42 \tau_\alpha^{-0.2}$.   
(c) FD plots for $\tau=9$, $\phi=0.819$ and different values of $q$. 
All wavevectors yield a unique value of the 
effective temperature $T_{\rm eff} \approx 7.2$ (dashed line).} 
\label{fig:k4pi_varPF}
\end{figure}

We compute the susceptibility $\chi_q(t)$ and correlations $F_s(q,t)$ 
associated to 
a modulated perturbation of wavevector $q$ in dense suspensions of 
self-propelled disks. We focus first on the parametric FD plots for fixed 
$q=4\pi$, $\tau=9$ and several packing fractions from $\phi=0.747$ 
(liquid regime) to $\phi=0.821$ (very close to the
non-equilibrium glass transition $\phi_c(\tau=9) \approx 0.823$). 
The results are displayed in Fig.~\ref{fig:k4pi_varPF}. 
Similar results are found for other values of the persistence time.

After a relatively 
short time corresponding to $F_s(q,t \approx \tau) \approx 0.85$ the FD plots  
for different packing fractions are well-described by 
straight lines with a slope that evolves slowly with $\phi$. 
In the density regime investigated, $T_{\rm eff}$ decreases from 
$T_{\rm eff} \approx 15$ to $T_{\rm eff} \approx 7$ over a density regime 
where the relaxation time increases by roughly seven orders of magnitude.
The short-time period corresponding to the non-thermal part of the FD plot
is a distinctive feature of active systems.
In aging glassy systems, this crossover corresponds to the 
two-step decay of correlation functions with an intermediate 
plateau regime. Our system presents a two-step decay of $F_s(q,t)$ as 
well~\cite{Berthier2013}, but the plateau value of the  
correlation function plays no role in the FD plots. 
Instead, deviations from equilibrium are controlled by the 
timescale of the self-propulsion mechanism, an effective 
temperature appearing when $t \gg \tau$. 
As similar trend was noted in mean-field calculations\cite{BerthierKurchan2013}.

From the data in Fig.~\ref{fig:k4pi_varPF}(a) we
extract the value of the effective temperature 
and present its evolution with density in Fig.~\ref{fig:k4pi_varPF}(b). 
To better appreciate the distance to the glass transition 
we simultaneously measure the structural relaxation time 
$\tau_\alpha$ of the system defined as $F_s(q_{\rm max},\tau_\alpha) = 0.2$, 
and represent $T_{\rm eff}$ as a function of $\tau_{\alpha}$; 
$q_{max}$ corresponds to the first peak of the static structure factor. 
These results show that $T_{\rm eff}$ decays as dynamical 
arrest is approached, and we estimate that it approaches 
an asymptotic value $T_{\rm eff} \approx 7.2$ as $\tau_\alpha \to \infty$, 
which clearly differs from the equilibrium value. 
Therefore we see that the persistence time $\tau$ not only 
shifts the location of the glass transition towards large densities, 
it also fixes the value of the effective temperature at the glass transition. 

The crucial prediction of the mean-field theory of 
glassy dynamics is the observable-independence of the effective temperature.
To test this, we compute 
$\chi_q(t)$ and $F_s(q,t)$ for several different wavevectors 
at fixed $\tau=9$ and $\phi=0.819$. 
Remarkably, the results shown in Fig.~\ref{fig:k4pi_varPF} 
are consistent with a {\it unique} effective temperature, as the FD plots 
for different $q$ are almost superimposed.
In particular, the difference with the data at lower 
density  in Fig.~\ref{fig:PF747_vark} is striking. Numerically, the 
trend is that the $q$-dependence observed in the fluid becomes 
less and less pronounced as the density is increased towards the glassy phase,
in perfect agreement with mean-field predictions.

\section{Conclusion}

The central result of the present work is the 
numerical confirmation in a finite dimensional model of self-propelled
particles that a collective effective temperature
emerges in active systems near a non-equilibrium glass transition. 
We have also shown that this 
physical picture does not apply in the active fluid 
at small and moderate densities, where only perturbations 
at large enough lengthscales appear effectively ``thermal'',
whereas finite lengthscales perturbations are not well described 
by an effective thermodynamical treatment.  

Physically, the key aspect of active glassy systems that 
allows a meaningful definition of the  
effective temperature, is a clear separation of timescales between 
microscopic dynamics at small timescales that is highly sensitive to details 
of the self-propulsion and the structural relaxation taking place 
at much larger timescales. It is the long-time 
sector for which the definition of an effective temperature 
appears meaningful. In addition, because the 
system undergoes a simultaneous 
kinetic freezing of all degrees of freedom near the glass
transition, the effective temperature becomes independent
of the probed lengthscales and is thus essentially shared 
by all of them.

\acknowledgments 

The research leading to these results has received funding
from the European Research Council under the European Union's Seventh
Framework Programme (FP7/2007-2013) / ERC Grant agreement No 306845. 

\bibliographystyle{unsrt}

\bibliography{FDT02}

\begin{thebibliography}{10}

\bibitem{ChandlerBook}
D.~Chandler.
\newblock {\em {Introduction to Modern Statistical Mechanics}}.
\newblock Oxford University Press, 1987.

\bibitem{Kurchan2005}
J.~Kurchan.
\newblock {\em Nature}, 433:222, 2005.

\bibitem{Angelini2011}
T.~E. Angelini, E.~Hannezo, X.~Trepat, M.~Marquez, J.~J. Fredberg, and D.~A.
  Weitz.
\newblock {\em PNAS}, 108(12):4714--9, 2011.

\bibitem{WuLibchaber2000}
X-L. Wu and A.~Libchaber.
\newblock {\em Phys. Rev. Lett.}, 84:3017, 2000.

\bibitem{Palacci2010}
J.~Palacci, C.~Cottin-Bizonne, C.~Ybert, and L.~Bocquet.
\newblock {\em Phys. Rev. Lett.}, 105(8):088304, 2010.

\bibitem{Theurkauff2012}
I.~Theurkauff, C.~Cottin-Bizonne, J.~Palacci, C.~Ybert, and L.~Bocquet.
\newblock {\em Phys. Rev. Lett.}, 108(26):268303, 2012.

\bibitem{Buttinoni2013}
I.~Buttinoni, J.~Bialk\'{e}, F.~K\"{u}mmel, H.~L\"{o}wen, C.~Bechinger, and
  T.~Speck.
\newblock {\em Phys. Rev. Lett.}, 110(23):238301, 2013.

\bibitem{Ginot2014}
F.~Ginot, I.~Theurkauff, D.~Levis, C.~Ybert, L.~Bocquet, L.~Berthier, and
  C.~Cottin-Bizonne.
\newblock {\em Phys. Rev. X}, 5:011004, 2015.

\bibitem{Deseigne2010}
J.~Deseigne, O.~Dauchot, and H.~Chat\'{e}.
\newblock {\em Phys. Rev. Lett.}, 105:098001, 2010.

\bibitem{ramaswamy}
V.~Narayan, S.~Ramaswamy, and N.~Menon.
\newblock {\em Science}, 317:105, 2007.

\bibitem{Marchetti2012a}
M.~C. Marchetti, J.~F. Joanny, S.~Ramaswamy, T.~B. Liverpool, J.~Prost, M.~Rao,
  and R.~Aditi~Simha.
\newblock {\em Rev. Mod. Phys.}, 85:1143, 2013.

\bibitem{VicsekReview}
T.~Vicsek and A.~Zafeiris.
\newblock {\em Physics Reports}, 517:71--140, 2012.

\bibitem{Loi2008}
D.~Loi, S.~Mossa, and L.~F. Cugliandolo.
\newblock {\em Phys. Rev. E}, 77:051111, 2008.

\bibitem{Wang2011a}
S.~Wang and P.~G. Wolynes.
\newblock {\em J. Chem. Phys.}, 135(5):051101, 2011.

\bibitem{Loi2011}
D.~Loi, S.~Mossa, and L.~F. Cugliandolo.
\newblock {\em Soft Matter}, 7:10193, 2011.

\bibitem{Ben-Isaac2011}
E.~Ben-Isaac, Y-K. Park, G.~Popescu, F.~Brown, N.~Gov, and Y.~Shokef.
\newblock {\em Phys. Rev. Lett.}, 106:238103, 2011.

\bibitem{Szamel2014a}
G.~Szamel.
\newblock {\em Phys. Rev. E}, 90(1):012111, 2014.

\bibitem{Suma2014}
A.~Suma, G.~Gonnella, G.~Laghezza, A.~Lamura, A.~Mossa, and L.~F. Cugliandolo.
\newblock {\em Phys. Rev. E}, 90(5):052130, 2014.

\bibitem{Takatori2014}
S.~C. Takatori, W.~Yan, and J.~F. Brady.
\newblock {\em Phys. Rev. Lett.}, 113(2):028103, 2014.

\bibitem{Solon2014}
A.~P. Solon, Y.~Fily, A.~Baskaran, M.~E. Cates, Y.~Kafri, M.~Kardar, and
  J.~Tailleur.
\newblock {\em arXiv: 1412.3952}, 2014.

\bibitem{Farage2015}
T.~F.~F. Farage, P.~Krinninger, and J.~M. Brader.
\newblock {\em Phys. Rev. E}, 91:042310, 2015.

\bibitem{Tailleur2008}
J.~Tailleur and M.~E. Cates.
\newblock {\em Phys. Rev. Lett}, 100:218103, 2008.

\bibitem{Speck2014}
J.~Bialk\'e, H.~L\"owen, and T.~Speck.
\newblock {\em arXiv:1412.4601}, 2014.

\bibitem{Cugliandolo1993}
L.~F. Cugliandolo and J.~Kurchan.
\newblock {\em Phys. Rev. Lett.}, 71:173, 1993.

\bibitem{Cugliandolo1994}
L.~F. Cugliandolo and J.~Kurchan.
\newblock {\em J. Phys. A: Math. Theor.}, 27:5749, 1994.

\bibitem{Crisanti2003}
A.~Crisanti and F.~Ritort.
\newblock {\em J. Phys. A: Math. Theor.}, 36:R181, 2003.

\bibitem{Cugliandolo2011}
L.~F Cugliandolo.
\newblock {\em J. Phys. A: Math. Theor.}, 44(48):483001, 2011.

\bibitem{Cugliandolo1997}
L.~F. Cugliandolo, J.~Kurchan, and L.~Peliti.
\newblock {\em Phys. Rev. E}, 55:3898, 1997.

\bibitem{Tailleur2009}
J.~Tailleur and M.~E. Cates.
\newblock {\em EPL}, 86:60002, 2009.

\bibitem{Maggi2014}
C.~Maggi, M.~Paoluzzi, N.~Pellicciotta, A.~Lepore, L.~Angelani, and
  R.~Di~Leonardo.
\newblock {\em Phys. Rev. Lett.}, 23(113):238303, 2014.

\bibitem{Berthier2013}
L.~Berthier and J.~Kurchan.
\newblock {\em Nature Physics}, 9(5):310--314, 2013.

\bibitem{Levis2014}
D.~Levis and L.~Berthier.
\newblock {\em Phys. Rev. E}, 89(6):062301, 2014.

\bibitem{Berthier2014}
L.~Berthier.
\newblock {\em Phys. Rev. Lett.}, 112(22):220602, 2014.

\bibitem{KrauthBook}
W.~Krauth.
\newblock {\em {Statistical Mechanics: Algorithms and Computations}}.
\newblock Oxford University Press, 2006.

\bibitem{Szamel2015}
G.~Szamel, E.~Flenner, and L.~Berthier.
\newblock {\em Phys. Rev. E}, 91:062304, 2015.

\bibitem{Donev2006}
A.~Donev, F.~H. Stillinger, and S.~Torquato.
\newblock {\em Phys. Rev. Lett.}, 96:0603183, 2006.

\bibitem{Berthier2002}
L.~Berthier and J-L. Barrat.
\newblock {\em J. Chem. Phys.}, 116(14):6228, 2002.

\bibitem{Sollich2002}
S.~Fielding and S.~Sollich.
\newblock {\em Phys. Rev. Lett.}, 88:050603, 2002.

\bibitem{Redner2013}
G.~S. Redner, M.~F. Hagan, and A.~Baskaran.
\newblock {\em Phys. Rev. Lett.}, 110:055701, 2013.

\bibitem{Fily2012}
Y.~Fily and M.~C. Marchetti.
\newblock {\em Phys. Rev. Lett.}, 108(23):235702, 2012.

\bibitem{BerthierKurchan2013}
L.~Berthier and J.~Kurchan.
\newblock {\em Nature Physics}, 9(1):310--314, 2013.

\bibitem{Ni2013}
R.~Ni, M.~A. {Cohen Stuart}, and M.~Dijkstra.
\newblock {\em Nature Comm.}, 4:2704, 2013.

\bibitem{Wysocki2014a}
A.~Wysocki, R.~G. Winkler, and G.~Gompper.
\newblock {\em EPL}, 105:48004, 2014.

\bibitem{Brader2014}
T.~F.~F. Farage and J.~M. Brader.
\newblock {\em arXiv:1403.0928}, 2014.

\bibitem{dasgupta}
R.~Mandal, P.~J. Bhuyan, M.~Rao, and C.~Dasgupta.
\newblock {\em arXiv:1412.1631}, 2014.

\end{thebibliography}

\end{document}